\documentclass[aps,prl,twocolumn,preprintnumbers,amsmath,amssymb,superscriptaddress]{revtex4} 
\usepackage{graphicx}
\usepackage{bm}
\usepackage[final]{pdfpages}

\begin{document}
\title{Observation of Anomalous Magnetic Moments in Superconducting Bi/Ni Bilayer}
\date{\today} 

\author{Junhua Wang} \thanks{These authors contribute equally to this work.} \affiliation{Beijing National Laboratory for Condensed Matter Physics, Institute of Physics, Chinese Academy of Sciences, Beijing 100190, People's Republic of China}
\author{Xinxin Gong} \thanks{These authors contribute equally to this work.} \affiliation{State Key Laboratory of Surface Physics and Department of Physics, Fudan University, Shanghai 200433, People's Republic of China}
\author{Guang Yang} \affiliation{Beijing National Laboratory for Condensed Matter Physics, Institute of Physics, Chinese Academy of Sciences, Beijing 100190, People's Republic of China}
\author{Zhaozheng Lyu} \affiliation{Beijing National Laboratory for Condensed Matter Physics, Institute of Physics, Chinese Academy of Sciences, Beijing 100190, People's Republic of China}
\author{Yuan Pang} \affiliation{Beijing National Laboratory for Condensed Matter Physics, Institute of Physics, Chinese Academy of Sciences, Beijing 100190, People's Republic of China}
\author{Guangtong Liu} \affiliation{Beijing National Laboratory for Condensed Matter Physics, Institute of Physics, Chinese Academy of Sciences, Beijing 100190, People's Republic of China}
\author{Zhongqing Ji} \affiliation{Beijing National Laboratory for Condensed Matter Physics, Institute of Physics, Chinese Academy of Sciences, Beijing 100190, People's Republic of China}
\author{Jie Fan} \affiliation{Beijing National Laboratory for Condensed Matter Physics, Institute of Physics, Chinese Academy of Sciences, Beijing 100190, People's Republic of China}
\author{Xiunian Jing} \affiliation{Beijing National Laboratory for Condensed Matter Physics, Institute of Physics, Chinese Academy of Sciences, Beijing 100190, People's Republic of China} \affiliation{Collaborative Innovation Center of Quantum Matter, Beijing 100871, People's Republic of China}
\author{Changli Yang} \affiliation{Beijing National Laboratory for Condensed Matter Physics, Institute of Physics, Chinese Academy of Sciences, Beijing 100190, People's Republic of China} \affiliation{Collaborative Innovation Center of Quantum Matter, Beijing 100871, People's Republic of China}
\author{Fanming Qu} \affiliation{Beijing National Laboratory for Condensed Matter Physics, Institute of Physics, Chinese Academy of Sciences, Beijing 100190, People's Republic of China}
\author{Xiaofeng Jin} \email[Corresponding authors. Email: ]{lilu@iphy.ac.cn, xfjin@fudan.edu.cn} \affiliation{State Key Laboratory of Surface Physics and Department of Physics, Fudan University, Shanghai 200433, People's Republic of China}
\author{Li Lu} \email[Corresponding authors. Email: ]{lilu@iphy.ac.cn, xfjin@fudan.edu.cn} \affiliation{Beijing National Laboratory for Condensed Matter Physics, Institute of Physics, Chinese Academy of Sciences, Beijing 100190, People's Republic of China} \affiliation{Collaborative Innovation Center of Quantum Matter, Beijing 100871, People's Republic of China} \affiliation{School of Physical Sciences, University of Chinese Academy of Sciences, Beijing 100190, People's Republic of China}

\begin{abstract}
  There have been continuous efforts in searching for unconventional superconductivity over the past five decades. Compared to the well-established $d$-wave superconductivity in cuprates, the existence of superconductivity with other high-angular-momentum pairing symmetries is less conclusive. Bi/Ni epitaxial bilayer is a potential unconventional superconductor with broken time reversal symmetry (TRS), for that it demonstrates superconductivity and ferromagnetism simultaneously at low temperatures. We employ a specially designed superconducting quantum interference device (SQUID) to detect, on the Bi/Ni bilayer, the orbital magnetic moment which is expected if the TRS is broken. An anomalous hysteretic magnetic response has been observed in the superconducting state, providing the evidence for the existence of chiral superconducting domains in the material.
\end{abstract}

\pacs{}

\maketitle

In searching for unconventional superconductivity with broken TRS, it was proposed that spin triplet pairing could be triggered by magnetic fluctuations in ferromagnetic metals \cite{Fay,Julian,Hattori}, with shared itinerant electrons responsible for both superconductivity and ferromagnetism \cite{Machida,Shopova}. The coexistence of superconductivity and ferromagnetism at low temperatures has indeed been reported in UGe$_2$ \cite{Saxena}, URhGe \cite{Aoki} and UCoGe \cite{Huy}. Spin triplet pairing was believed also to be able to occur via proximity effect at the interface between $s$-wave superconductor and ferromagnetic metal \cite{Bergeret,Kadigrobov}. The existence of a supercurrent across a NbTiN-CrO$_2$-NbTiN junction was regarded as the evidence \cite{Keizer}. The most extensively studied candidate of $p$-wave superconductor is probably Sr$_2$RuO$_4$ \cite{Sigrist1991,Mackenzie2003,Maeno2012}, with which anomalous responses in phase sensitive experiments \cite{Nelson2004,Kidwingira2006,Bouhon2010} and the appearance of half flux quantum \cite{JJang} have been observed. However, the existence of domain-edge current, a signature for chiral superconductivity due to broken TRS, has not yet been proven \cite{Bjrnsson,Kirtley,Hicks}.

As a potential unconventional superconductor, Bi/Ni bilayer film was firstly studied by Moodera and Meservey \cite{Moodera}, and later by LeClair and coworkers \cite{LeClair}. Superconductivity was observed to coexist with ferromagnetism below $T_{\rm c}=4.2$ K, and was thought to originate from a possible \emph{fcc}-structured Bi on \emph{fcc}-structured Ni \cite{LeClair}. Recently, Gong and coworkers further revealed the coexistence of superconductivity and ferromagnetism in an epitaxial form of Bi/Ni bilayer which contains only ordinary rhombohedra instead of \emph{fcc}-structured Bi \cite{Gong}. Since neither rhombohedra Bi nor \emph{fcc} Ni is superconducting individually at the temperatures explored, the superconductivity is possibly triggered by magnetic fluctuation at the interface and hence likely to be unconventional \cite{Fay,Gong1}.

If the superconductivity in Bi/Ni bilayer is indeed unconventional and contains chiral superconducting domains due to broken TRS, there will be observable out-of-plane magnetic moments at the edges (Fig. 1\textbf{a}) due to the positional mismatch (hence the incomplete cancellation) between the outmost domain-edge current and the inward Meissner screening supercurrent \cite{Sigrist1989}. In this work, we have constructed SQUIDs in such a way that a portion of the interference loop is the Bi/Ni bilayer itself (Fig. 2\textbf{a}). This configuration is most sensitive for exploring edge magnetization --- any variation in edge magnetization will directly change the total flux in the SQUID loop (Fig. 1\textbf{b}). We have indeed observed an anomalous hysteretic interference pattern on such SQUIDs. We attribute the hysteresis to the motion of chiral superconducting domains in the bilayer.

The SQUIDs we used contain two half-rings, as shown in Fig. 2\textbf{a}. One half-ring is etched off from the Bi/Ni bilayer, with a width of 1 $\mu$m and inner radii listed in Tab. \uppercase\expandafter{\romannumeral1}. The Bi/Ni bilayer contains 2 nm-thick \emph{fcc}-structured Ni and 30 nm-thick rhombohedra-structured Bi, grown sequentially on MgO substrate via molecular beam epitaxy \cite{Gong}. The other half-ring is made of superconducting Pb film, which is 50 nm thick and with a width of 1.4 $\mu$m. Au film pads of 30 nm thick are sandwiched between the two half-rings. The strong superconducting proximity effect between Pb and Au film guarantees Josephson couplings between the two half-rings up to a distance of $>0.5$ $\mu$m. We note that the use of a Pb half-ring is necessary, because the superconducting proximity effect between Bi/Ni and Au is not strong enough to mediate Josephson supercurrents through Bi/Ni-Au-Bi/Ni junctions.

\begin{figure}
\includegraphics[width=1 \linewidth]{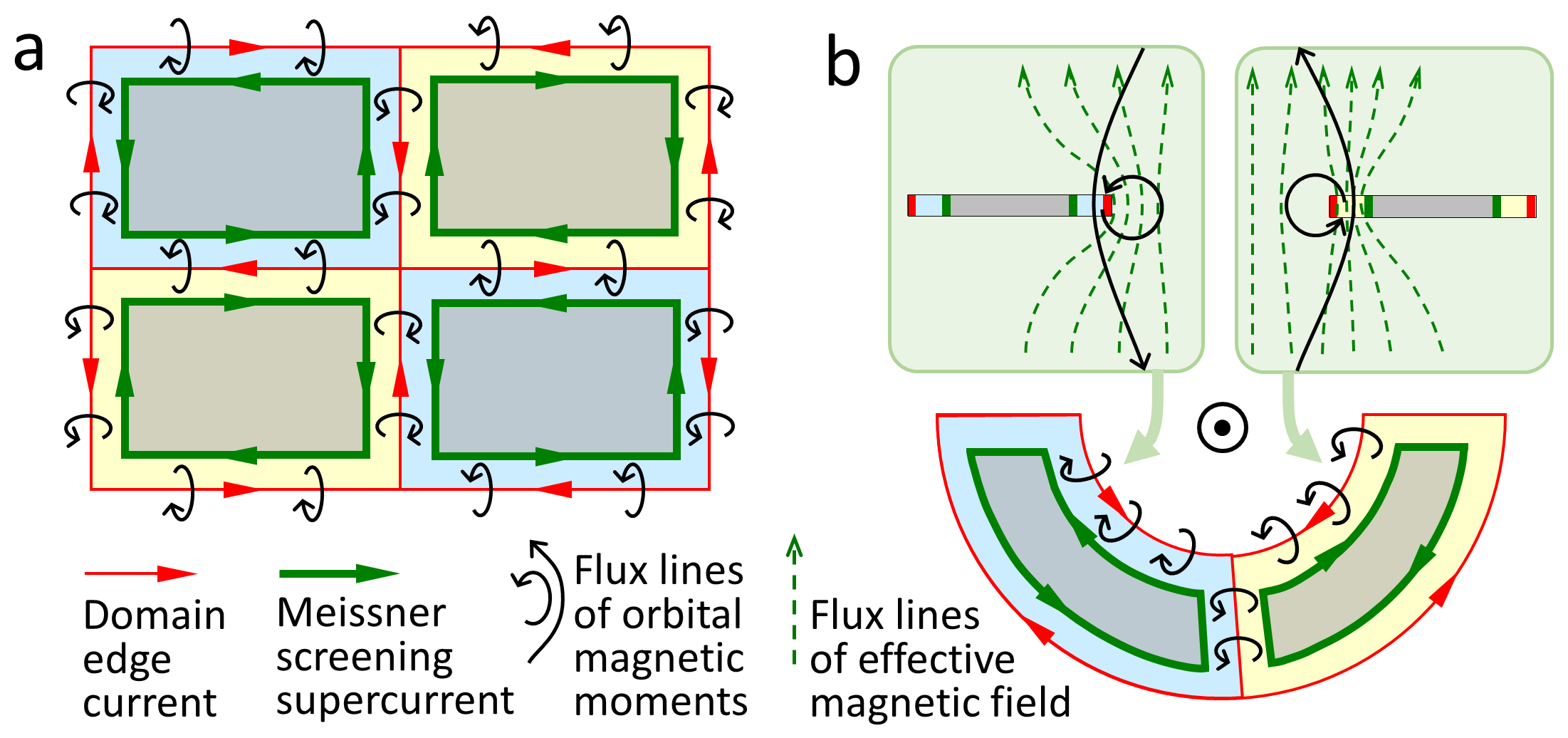}
\caption{\label{fig:fig1} {(color online) (\textbf{a}) A cartoon picture of chiral superconducting domains. The orbital magnetic moments of Cooper pairs are screened inside the paths of Meissner screening supercurrent (i.e., inside the gray areas), but unscreened near the edges. (\textbf{b}) (Lower panel) Illustration of a Bi/Ni half-ring containing two chiral superconducting domains of opposite chirality. (Upper panel) Cross-section views for the flux lines of effective magnetic field at the inner edges of the Bi/Ni half-ring.}}
\end{figure}

\begin{figure}
\includegraphics[width=1 \linewidth]{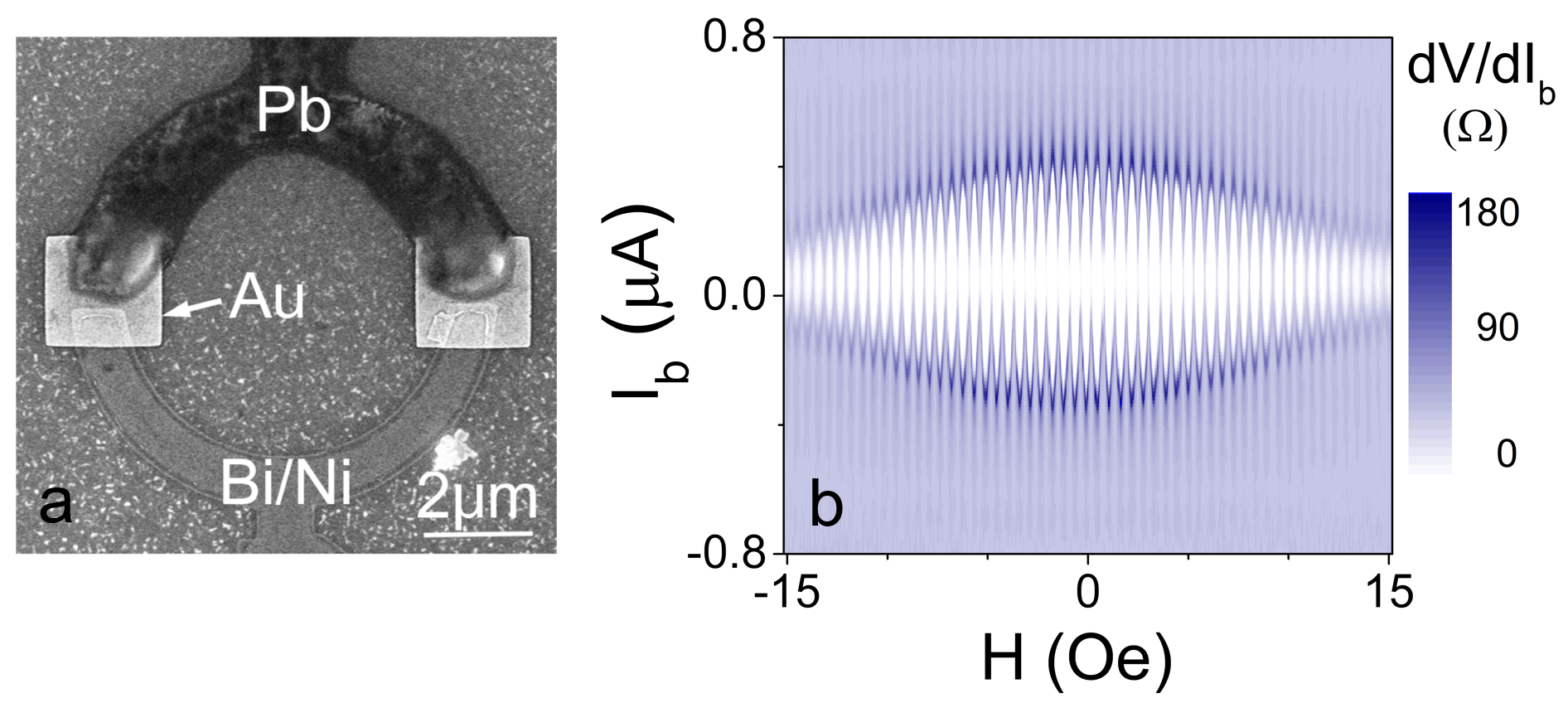}
\caption{\label{fig:fig2} {(color online) (\textbf{a}) Scanning electron microscope image of device \#1. The upper half-ring is made of Pb film, and the lower half-ring is made of Bi/Ni epitaxial bilayer. They are connected by two square-like Au films through superconducting proximity effect at low temperatures. The whitish structures at the edge of the Bi/Ni half-ring are the residual PMMA which is difficult to lift off after reactive ion etching. (\textbf{b}) 2D plot of differential resistance $dV/dI_{\rm b}$ as a function of out-of-plane magnetic field and bias current at $T=30$ mK.}}
\end{figure}

Six such SQUIDs were investigated at a temperature of $\sim$ 30 mK provided by dilution refrigerators. The differential resistance $dV/dI_{\rm b}$ of the devices was measured as a function of bias current $I_{\rm b}$ and out-of-plane magnetic fields $H$ by using lock-in amplifier techniques. Roughly periodic interference patterns were observed for all devices. A typical such pattern in unidirectional magnetic field sweepings is shown in Fig. 2\textbf{b}. The period of the oscillations is $\Delta B=0.55 \pm$0.03 Oe, which is in excellent agreement with the loop area of the device, $A_{\rm loop}=39.3 \mu$m$^2$, via the relation $A_{\rm loop}\Delta B=\phi_0$ (where $\phi_0$ is the flux quantum).

\begin{table*}
  \centering
  \caption{\label{table1} A list of the parameters of six devices investigated.}
 \begin{ruledtabular}
 \begin{tabular}{ccccccc}
  Device & Distance$^*$  & Inner Radius  & Loop Area  & Calculated Period  & Measured Period  & 2$\delta_0/2\pi$ \\
   &($\mu$m) & ($\mu$m) & (${\mu}m^2$)& (Oe)& (Oe) &  \\
  \hline
  $\#$1 & 0.4 & 3.0 & 39.3 & 0.53 & 0.55$\pm$0.03 & 2.89$\pm$0.01\\
  $\#$2 & 0.5 & 1.0 & 6.2 & 3.34 & 3.37$\pm$0.06 & 0.54$\pm$0.04\\
  $\#$3 & 0.2 & 1.0 & 6.1 & 3.39 & 3.13$\pm$0.08 & 0.70$\pm$0.04\\
  $\#$4 & -0.4 & 1.0 & 5.6 & 3.69 & 3.5$\pm$0.2 & 0.60$\pm$0.07\\
  $\#$5 & -0.4 & 2.0 & 17.6 & 1.18 & 1.12$\pm$0.09 & 1.7$\pm$0.2\\
  $\#$6 & -0.4 & 3.0 & 35.9 & 0.59 & 0.58$\pm$0.03 & 2.7$\pm$0.2\\
\end{tabular}
\end{ruledtabular}
\begin{flushleft}
  $^*$ The lateral distance between Pb and Bi/Ni at the junctions. A negative value denotes the overlap of the two half-rings.
\end{flushleft}
\end{table*}

However, we noticed that the device enters into a transient state whenever the sweep direction of magnetic field is reversed. The oscillations are compressed in this transient state. Figure 3\textbf{a} shows the $dV/dI_{\rm b}$ oscillations in two sweep circles, measured on device \#1 at $I_{\rm b}=0$. Let us firstly look at the green trace which was measured by sweeping field from the right to the left. The oscillations are compressed at the beginning, then gradually recover to the normal periods (entering into a stable state). The range of the transient state, $H_{\rm t}$, is marked by the light-green color bar at the top. Once the sweep direction is reversed from the left to the right ({\it i.e.}, the red trace), the oscillations are compressed again at the beginning, then recover to the normal periods again.

\begin{figure*}
\includegraphics[width=0.8 \linewidth]{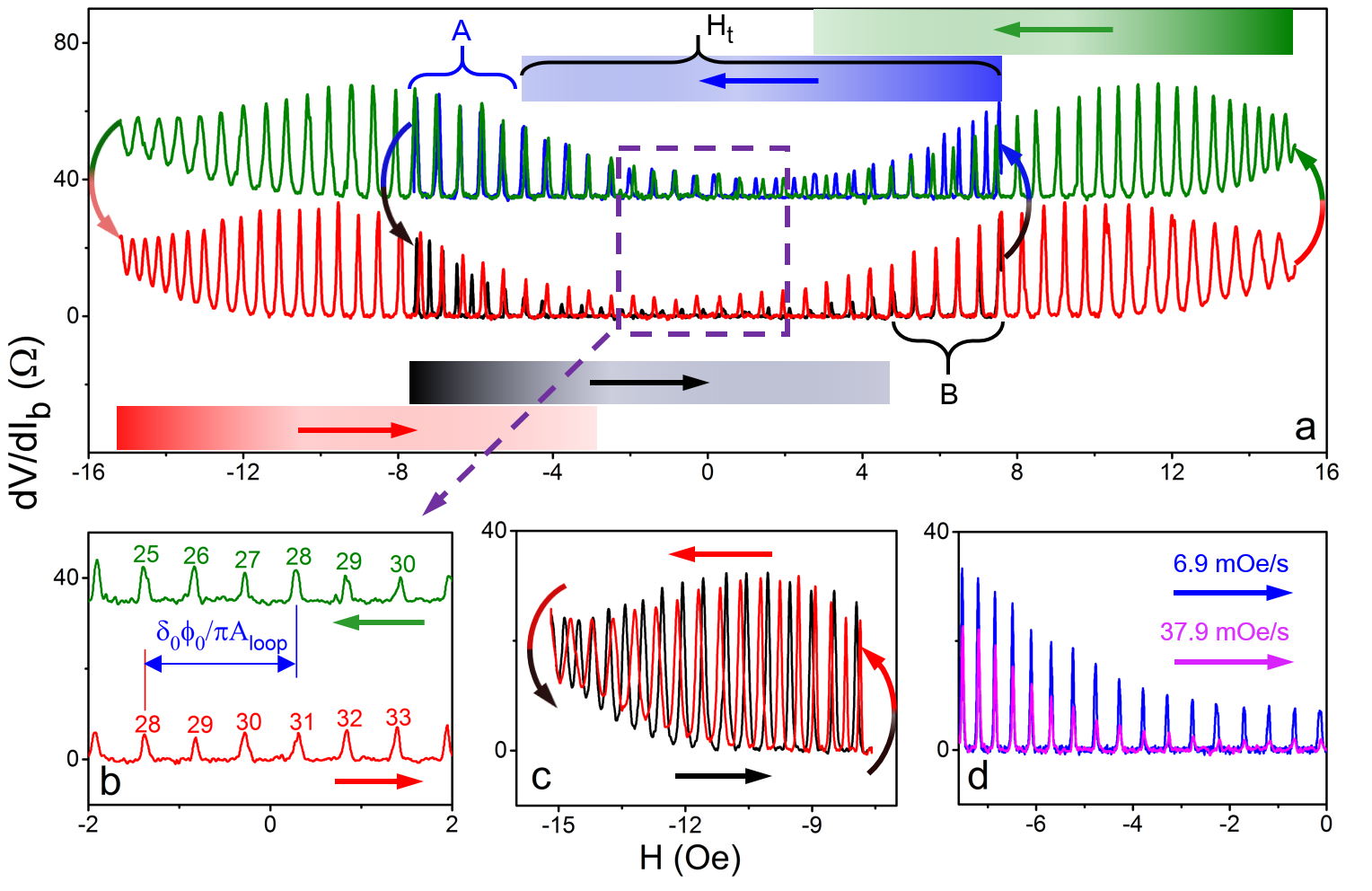}
\caption{\label{fig:fig3} {(color online) Anomalous hysteresis on the superconducting interference pattern of device \#1, measured at $I_{\rm b}=0$ and with an ac excitation current of 0.01 $\mu$A at $T=30$ mK. (\textbf{a}) Traces of $dV/dI_{\rm b}$ taken in two sweep circles with different magnetic field ranges. The traces with green and blue colors are shifted vertically for clarity. The arrows indicate the sweep directions, and bars of length $H_{\rm t}$ denote the transient states, with their contrast representing the extent of compression in periods. Regions A and B stand for the stable states for the blue and the black traces, respectively. (\textbf{b}) The green and the red traces in their stable states. The peak indexes are counted from the left side of the sweep circle in (a). (\textbf{c}) A hysteretic circle in negative magnetic fields, showing that it is not necessary to surround the zero field. (\textbf{d}) The peak positions measured in the same sweep direction but with different sweep rates overlap with each other.}}
\end{figure*}

In Fig. 3\textbf{a} we also show a smaller sweep circle containing the blue trace and the black trace. Similar transient states exist, with about the same length of $H_{\rm t}$ compared to that on the green and the red traces. After having passed the transient states and entered into the stable state, {\it e.g.}, in region A for negative field sweep direction and in region B for positive field sweep direction, the peak positions on traces of same sweep directions overlap with each other.

Due to the compression in the transient states, however, the peaks with same indexes but on traces of opposite sweep directions are shifted with each other, as shown in Fig. 3\textbf{b}. Here we note that the peaks on the two traces of a closing circle are one-to-one correlated, and are indexed from the left to the right. The shift is ``advanced" with regard to the field sweep direction, such that a given indexed peak occurs before reaching its position on the trace of opposite sweep direction.

There are two additional features to be mentioned. One is that the transient state is only related to the reverse of the field sweep direction rather than the field strength. For example, Figure 3\textbf{c} shows a hysteretic circle in entirely negative magnetic fields. The other is that the peak positions in both the transient state and the stable state are independent of the field sweep rate. This can be readily seen from Fig. 3\textbf{d}. More data taken at different sweep rates can be found in the supplementary materials \cite{supplementary_materials}.

The observed anomalous hysteresis, appearing to be ``advanced" and occurring at any magnetic field once the sweep direction is reversed, is very different from the ordinary hysteresis caused by ferromagnetism or flux pinning in superconductors. It should be noted that there is no flux pinning in our devices, since the maximal out-of-plane magnetic field is kept well below 20 Oe during the measurements, which is insufficient to create a vortex on the Bi/Ni strip of 1 $\mu$m in width. The anomalous hysteresis is also not likely caused by the instrumental delay during data acquisition. In fact, the hysteresis appears to be unchanged even if the sweep rate is varied by three orders of magnitude \cite{supplementary_materials}. In addition, the SQUIDs are not in the hysteretic regime, since the SQUID screening parameter $\beta_{\rm e}=2{\pi}LI_{\rm c}/\phi_0$ is much smaller than 1 (where the inductance of the SQUID loop $L\approx 6$ pH, the critical supercurrent $I_{\rm c}<1$ $\mu$A) \cite{Barone}. All these arguments are well supported by our control experiment on conventional Pb-Au-Pb SQUIDs, in which no hysteresis can be recognized \cite{supplementary_materials}.

\begin{figure}
\includegraphics[width=0.95 \linewidth]{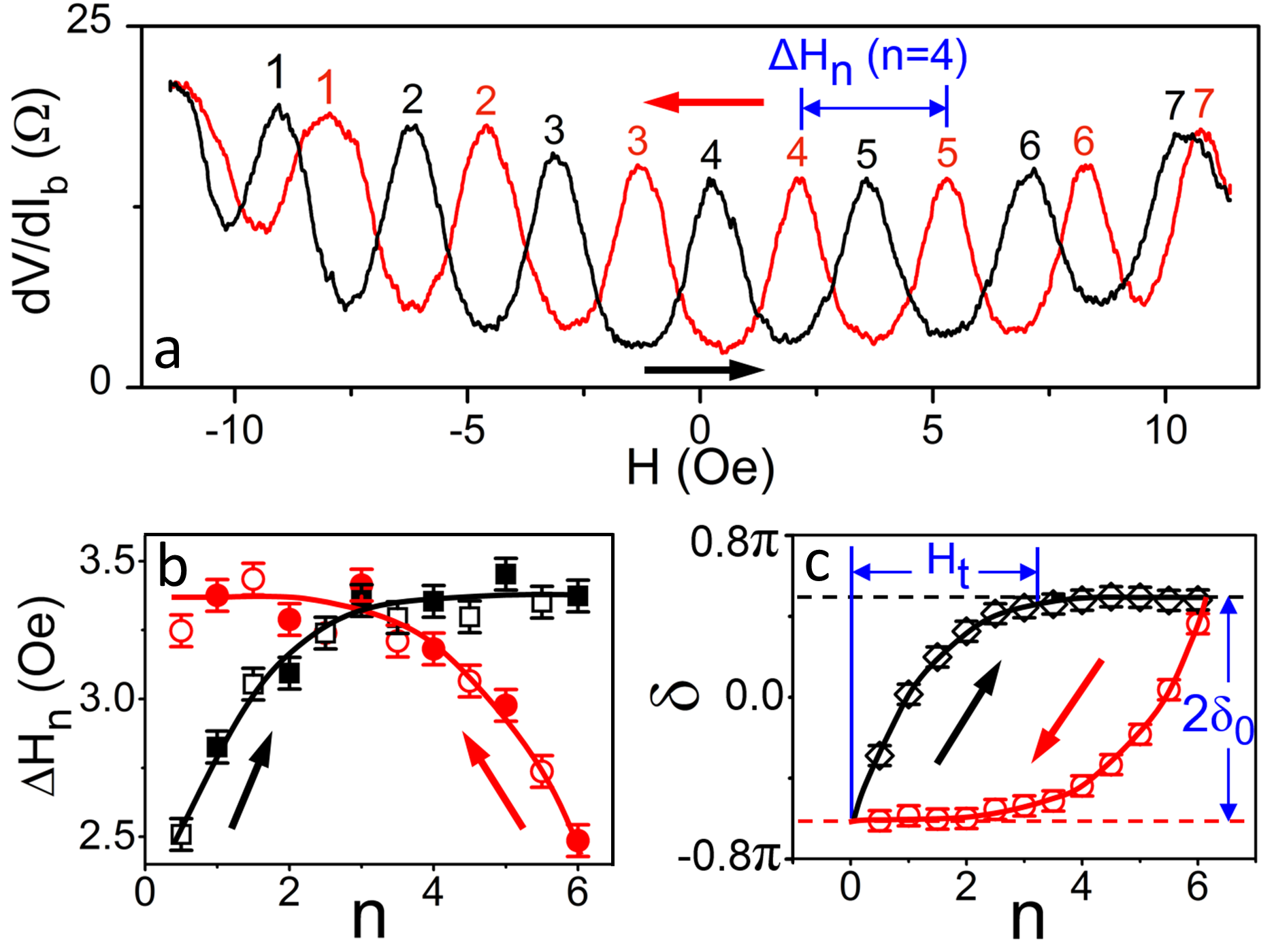}
\caption{\label{fig:fig4} {(color online) (\textbf{a}) Anomalous hysteresis of $dV/dI_{\rm b}$ measured on device \#2, with $I_{\rm b}=0.5$ $\mu$A and an ac excitation current of 0.05 $\mu$A at $T=30$ mK. The peaks are labeled with index $n$ starting from the left side of the circle. (\textbf{b}) The local period $\Delta H_{\rm n}$ between neighboring peaks (valleys) of indexes $n$ ($n-0.5$) and $n+1$ ($n+0.5$), is plotted against $n$ as solid (open) symbols. (\textbf{c}) The accumulated anomalous phase shift $\delta$. After having passed the transient states, $\delta$ reaches the saturated values as represented by the dashed black line and the dashed red line.}}
\end{figure}

To quantitatively analyze the hysteretic behavior, we collected a hysteresis circle of $dV/dI_{\rm b}$ on device \#2 at $I_{\rm b}=0.5$ $\mu$A. The result is shown in Fig. 4\textbf{a}. The oscillations of $dV/dI_{\rm b}$ are caused by the oscillations of $I_{\rm c}$, the critical supercurrent of the SQUID which has the form of \cite{Barone}:
\begin{equation}\label{equation1}
I_{\rm c} \propto \cos[2\pi\frac{(H-H_0)A_{\rm loop}}{\phi_0}+\delta]
\end{equation}
where $H_0$ is a small constant caused by, {\it e.g.}, the earth magnetic field. $\delta$ is the accumulated anomalous phase shift counted from the point where the sweep direction is reversed.

The phase changes between neighboring $dV/dI_{\rm b}$ peaks must satisfy:
\begin{equation}\label{equation2}
2\pi\frac{{\Delta}H_{\rm n}A_{\rm loop}}{\phi_{0}}+\Delta\delta_{\rm n}=2\pi
\end{equation}
where $\Delta H_{\rm n}$ is the field interval between the neighboring $dV/dI_{\rm b}$ peaks of indexes $n$ and $n+1$, as illustrated in Fig. 4\textbf{a} and plotted in Fig. 4\textbf{b}, and $\Delta\delta_{\rm n}$ is the difference of $\delta$ between the neighboring peaks of indexes $n$ and $n+1$.

By using Eq. 2 and the data of $\Delta H_{\rm n}$ shown in Fig. 4\textbf{b}, $\Delta\delta_{\rm n}$ can be calculated. The accumulated anomalous phase shift $\delta$ can be further obtained by adding up $\Delta\delta_{\rm n}$. The results are shown in Fig. 4\textbf{c}. It can be seen that, after the device having passed the transient states and reached the stable states, the accumulated anomalous phase shift $\delta$ saturates to $\delta_0$ or $-\delta_0$ as represented by the two dashed lines in Fig. 4\textbf{c}. Once the sweep direction is reversed, $\delta$ immediately starts to chase the other stable state, resulting in the observed ``advanced" nature.

Such an anomalous hysteresis is in sharp contrast to that of ferromagnetism. Therefore, it should be irrelevant to the ferromagnetic moment of the Ni layer. To further confirm this argument, we applied a magnetic field up to 1000 Oe in the in-plane direction of the bilayer, to drive the Ni layer into a single domain and to fully fix the direction of its ferromagnetic moment. According to earlier studies \cite{Gong}, the ferromagnetic moment of the Ni layer becomes saturated above 100 Oe. We find that the amount of anomalous phase shift between the two saturated states in out-of-plane magnetic field sweepings, 2$\delta_0$, keeps unchanged with varying in-plane magnetic fields. The results are shown in Figs. 5\textbf{a} to \textbf{e}. It confirms that the anomalous hysteresis is indeed unrelated to the ferromagnetic moment of the itinerant electrons in Ni.

\begin{figure}
\includegraphics[width=1 \linewidth]{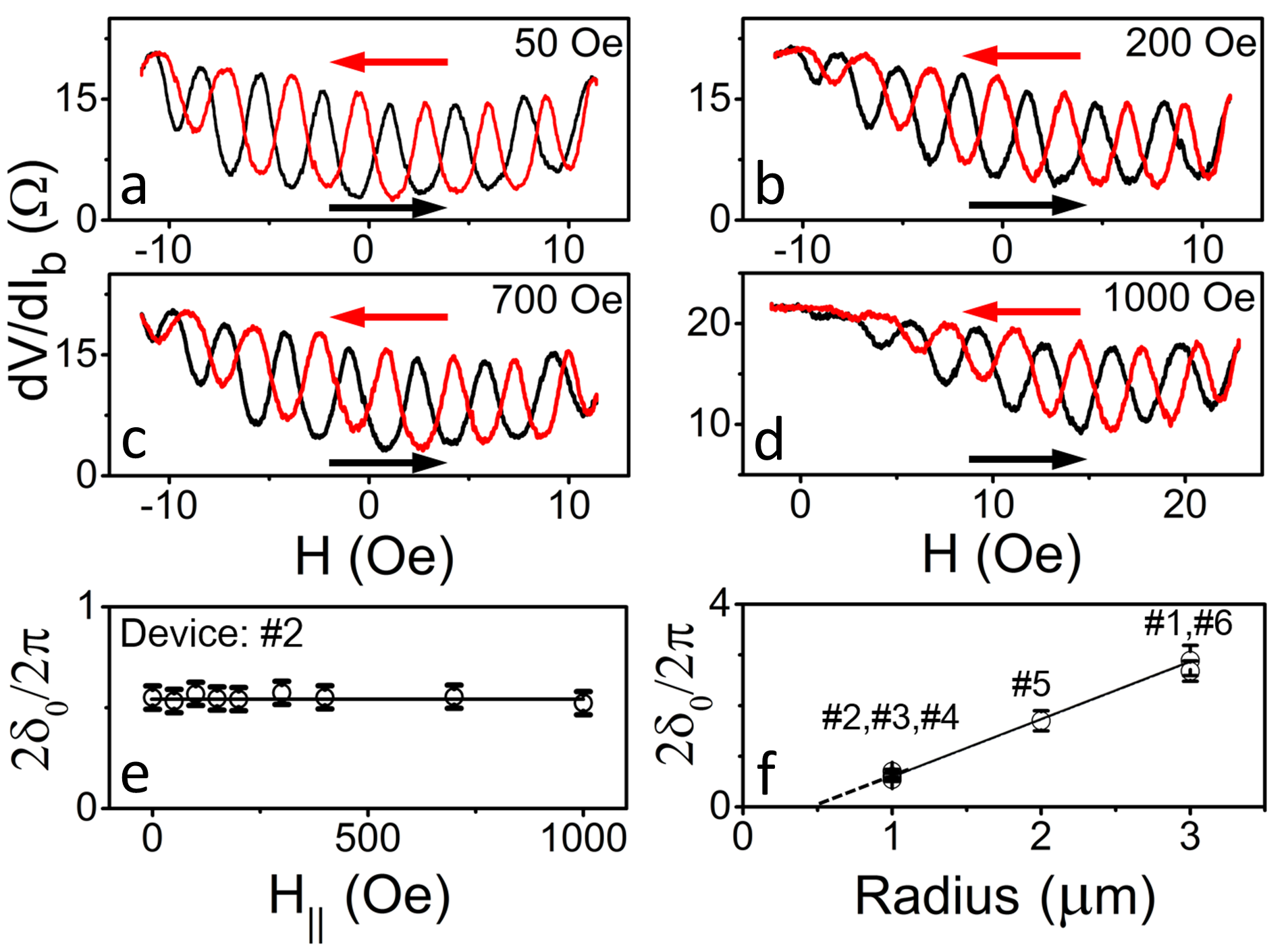}
\caption{\label{fig:fig5} {(color online) (\textbf{a}, \textbf{b}, \textbf{c} and \textbf{d}) Hysteresis of $dV/dI_{\rm b}$ measured on device \#2 at in-plane magnetic fields of $H_{\|}$ = 50, 200, 700, and 1000 Oe, respectively. The horizontal shift of the envelope is mainly caused by the out-of-plane component of the magnetic field. (\textbf{e}) $H_{\|}$-dependence of 2$\delta_0/2\pi$, the amount of hysteresis. (\textbf{f}) 2$\delta_0/2\pi$ of six devices plotted as a function of the inner radius of the Bi/Ni bilayer.}}
\end{figure}

A similar anomalous hysteretic behavior was observed in Sr$_2$RuO$_4$-Cu-Pb Josephson junctions by Kidwingira and coworkers, and was interpreted as the motion of superconducting domain walls \cite{Kidwingira2006,Bouhon2010}. We find that the same scenario is applicable to the phenomenon observed here. After assuming that there exist superconducting domains in the Bi/Ni bilayer, the observed hysteretic behavior can be ready to understand. The black and red horizontal lines in Fig. 4\textbf{c} represent two saturated states in which all flipable domains have already been aligned to the same direction. The curved parts of the traces in Fig. 4\textbf{c} correspond to the transient states within which the domain walls are moving to chase the saturated states.

We find that the length of the transient state, $H_{\rm t}$, keeps unchanged even if the field sweep rate is varied by several orders of magnitude. It indicates that the anomalous magnetic moment is able to respond very quickly to the change of applied magnetic field, and the response must also be self-controlled not to overshoot at any sweep rate. To explain this peculiar feature, we need to introduce two time scales $\tau_{\rm edge}$ and $\tau_{\rm Meissner}$. They are the characteristic time scales at which the domain edge current and the Meissner screening supercurrent circulating the entire Bi/Ni film respond to the change of applied magnetic field, respectively. Because the area varied by domain wall motion is smaller than the total area of the domains, we would expect that the corresponding inductance for these two currents are difference, leading to $\tau_{\rm edge} < \tau_{\rm Meissner}$. Once the applied magnetic field is changed, the domain walls have a time window of $\tau_{\rm Meissner}$ to move. After $\tau_{\rm Meissner}$, the Meissner screening supercurrent responds, so that the change of magnetic field is compensated and thus the motion of domain wall stops. This is why the hysteresis takes place whenever the field sweep direction is reversed, in spite of the strength of the field (see Fig. 3\textbf{c} for an example).

It is known that the domain-edge current flows at a characteristic distance of $\xi$ near the edge, while the Meissner screening supercurrent is distributed in a length scale of $\lambda_{\rm L}$ near the edge, where $\xi$ and $\lambda_{\rm L}$ are the coherence length and the penetration depth of the superconductivity in Bi/Ni bilayer, respectively. Usually $\xi<\lambda_{\rm L}$, namely these two current paths do not fully overlap with each other in space. Therefore, the orbital magnetic moments of the Cooper pairs near the edges are not fully screened by the Meissner screening supercurrents \cite{Sigrist1989}. These moments at the inner edge of the Bi/Ni half-ring contribute an anomalous flux to the SQUID loop in addition to the flux of applied magnetic field. A cartoon illustration of the mechanism is shown in Fig. 1. The motion of domain wall varies the anomalous flux in a hysteretic way, leading to the anomalous interference.

The accumulated anomalous phase shift reaches $\pm\delta_0$ when the inner edge of the Bi/Ni half-ring becomes fully polarized, i.e., becomes single-domained. $\delta_0$ is proportional to the total orbital magnetic moments of the Cooper pairs in the area $S$ between the two current paths at the inner edge of the Bi/Ni half-ring. Here $S=[\pi(R+\lambda_{\rm L})^2-\pi(R+\xi)^2]/2 \approx \pi(\lambda_{\rm L}-\xi) R$, where $R$ is the inner radius of the Bi/Ni half-ring, and $\lambda_{\rm L}-\xi$ is of the order of magnitude a hundred nanometers or so. This expression explains the linear radius dependence of $\delta_0$ shown in Fig. 5\textbf{f}. The intercept of the linear dependence at the abscissa in Fig. 5\textbf{f}, $\sim 0.5$ $\mu$m, might reflect the typical size of the superconducting domains --- below this size the Bi/Ni half-ring becomes single-domained, so that the domain edge overlaps with the ring edge and is pinned by the ring edge.

To summerize, we have employed a very sensitive device configuration to study the out-of-plane edge magnetization of the Bi/Ni bilayer in its superconducting state. An anomalous hysteretic behavior irrelevant to the ferromagnetism of Bi/Ni bilayer has been observed. We attribute the anomalous hysteresis to the motion of chiral superconducting domains in the bilayer. Our method could be applied to identify chiral superconductivity in other materials.


\noindent\textbf{Acknowledgments} We would like to thank Junya Feng for fruitful discussions. This work was supported by the MOST grants 2016YFA0300601, 2015CB921402, 2009CB929101 and 2011CB921702, by NSFC grants 91221203, 11174340, 11174357, 91421303, 11527806, 11434003, 11374057 and 11421404, and by the Strategic Priority Research Program B of the Chinese Academy of Sciences grant No. XDB07010100.

\begin{widetext}


\section*{Supplementary material}

# Supplementary Materials for:
# "Observation of Anomalous Magnetic Moments in Superconducting Bi/Ni Bilayer"

Junhua Wang,[1*] Xinxin Gong,[2*] Guang Yang,[1] Zhaozheng Lyu,[1] Yuan Pang,[1]
Guangtong Liu,[1] Zhongqing Ji,[1] Jie Fan,[1] Xiunian Jing,[1,3]
Changli Yang,[1,3] Fanming Qu,[1] Xiaofeng Jin,[2†] Li Lu[1,3,4†]

[1] Beijing National Laboratory for Condensed Matter Physics, and Institute of Physics,
Chinese Academy of Sciences, Beijing 100190, People's Republic of China

[2] State Key Laboratory of Surface Physics and Department of Physics, Fudan University,
Shanghai 200433, People's Republic of China

[3] Collaborative Innovation Center of Quantum Matter,
Beijing 100871, People's Republic of China

[4] School of Physical Sciences, University of Chinese Academy of Sciences,
Beijing 100190, People's Republic of China

*These authors contribute equally to this work.
†To whom correspondence should be addressed; E-mail: lilu@iphy.ac.cn, xfjin@fudan.edu.cn

# 1 Contents

**1. Estimation of the effective loop area of SQUID #1 in the main text**

**2. Influence of sweeping speed on the anomalous hysteretic behavior**

**3. Control experiment on SQUIDs with Pb-Au-Pb junctions**

# 1. Estimation of the effective loop area of SQUID #1 in the main text

In the presence of flux compression, we estimate the effective loop area of SQUID device #1 in the main text as follows (*1*):

$$\begin{aligned} Loop\ Area &= \pi R'_{outer}R'_{inner}/2 + \pi R_{outer}R_{inner}/2 + dw \\ &= \pi \times 2.8 \times 4.2/2 + \pi \times 3 \times 4/2 + 0.4 \times 5\ \mu m^2 \\ &= 39.3\ \mu m^2 \end{aligned}$$

As shown in Fig. S1, $R'_{outer}$ and $R'_{inner}$ are the outer and inner radii of the Pb half-ring, respectively, and $R_{outer}$ and $R_{inner}$ are the outer and inner radii of the Bi/Ni half-ring. $d$ is the lateral distance between the Pb film and the Bi/Ni bilayer, and $w$ is the separation between two square like Au films. We find that the superconducting proximity effect between Pb half-ring and Au film is very strong at 30 mK, so that the weakest parts that limit the Josephson supercurrent are likely located at the interface between superconducting Au and Bi/Ni bilayer. On the other hand, the induced superconductivity in An pad is not strong enough to exclude the magnetic flux, so that the effective SQUID loop is formed via the inner edges of the Au pads (i.e., the Au pads contribute to an additional area of $dw$).

# 2. Influence of sweeping speed on the anomalous hysteretic behavior

To check the influence of sweeping speed of magnetic field on the anomalous hysteresis, we varied the sweeping speed from 0.015 Oe/s to 7.59 Oe/s, and the hysteresis stays unchanged as shown in Fig. S2 (data taken on device #3 mentioned in the main text as an example). Therefore, the instrumental delay of data acquisition should not be responsible for the anomalous hysteresis, and the intrinsic time scale of this hysteresis must be much shorter than the time scale of

our measurement. In other words, the redistribution of superconducting domains must happen in a very short time and we can not "see" this process via our low-frequency measurement.

## 3. Control experiment on SQUIDs with Pb-Au-Pb junctions

The control experiment on SQUIDs consisting of Pb-Au-Pb junctions was carried out in the same dilution refrigerator after the measurement on the devices in the main text. No sign of hysteresis can be resolved, as shown in Fig. S3. Therefore, the anomalous hysteresis observed in the main text is unique to the Bi/Ni bilayer, but not because of the contribution from our measurement system such as residual flux pining in the magnet.

## References and Notes

1. Yuan Pang, Junhua Wang, Zhaozheng Lyu, Guang Yang, Jie Fan, Guangtong Liu, Zhongqing Ji, Xiunian Jing, Changli Yang, Li Lu (2016), "Observation of gap-closing in single Josephson junctions constructed on Bi2Te3 surface", arXiv:1603.04540.

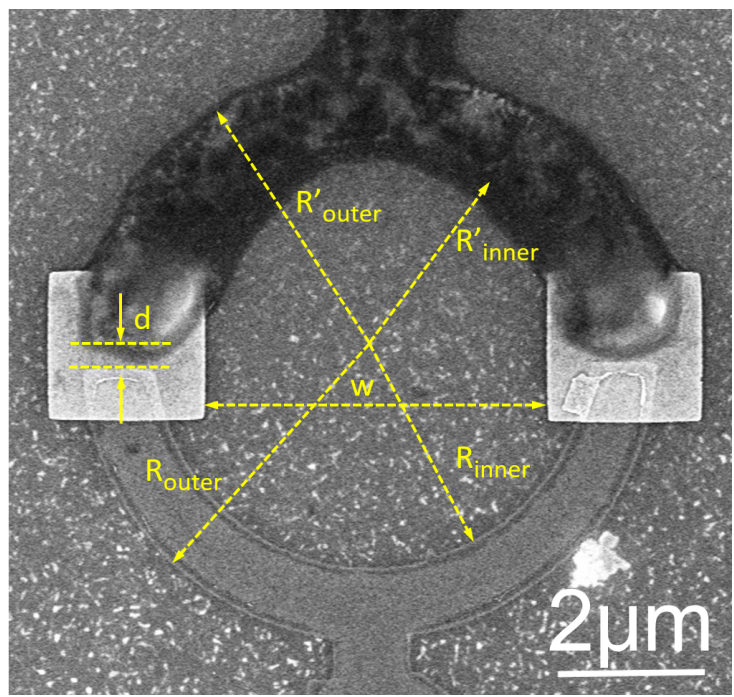

Fig. S1. Scanning electron microscope image of device #1 in the main text with the parameters indicated.

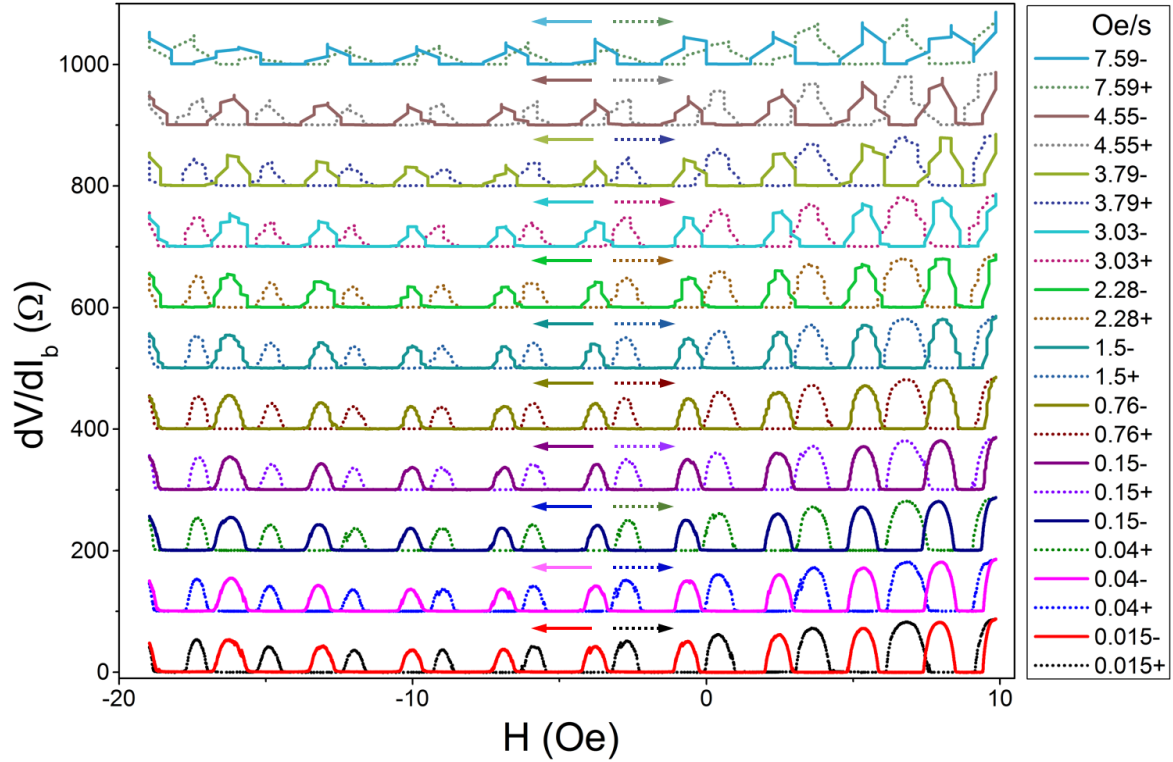

Fig. S2. The differential resistance $dV/dI_b$ of device #3 in the main text as a function of magnetic field. The data are taken at different field sweep rates from 0.015 Oe/s to 7.59 Oe/s (curves are offset for clarity). There is a delay of 0.1 s in data acquisition while sweeping magnetic field. The excitation current is 0.1 $\mu$A and the temperature is 30 mK. The solid and dotted curves correspond to backward and forward sweeps, respectively, as marked by the solid and dotted arrows.

5

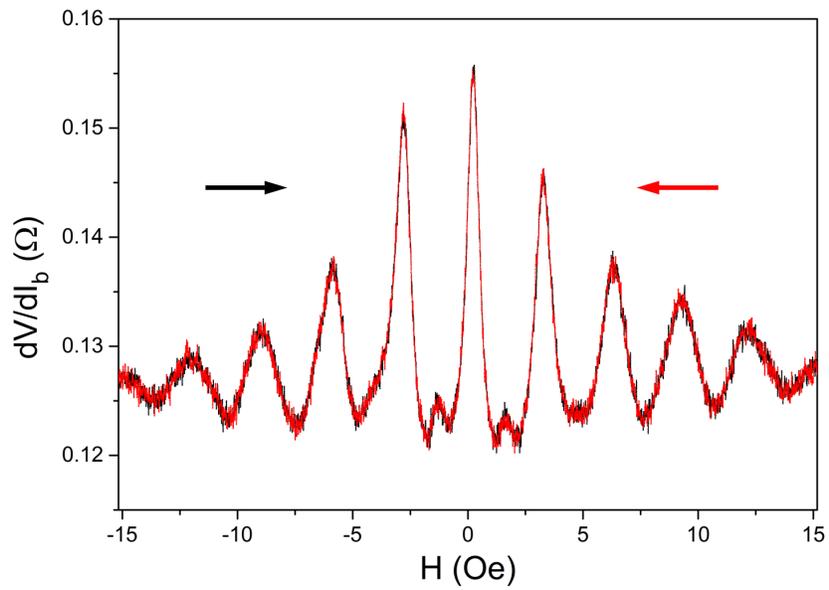

Fig. S3. The interference oscillation of a SQUID with Pb-Au-Pb junctions. The differential resistance $dV/dI_b$ is measured as a function of out-of-plane magnetic field with a bias current of $I_b = 200 \mu A$ and an ac excitation current of 5 $\mu$A, at temperature T=1 K. The field sweep rate is $\pm$ 0.0076 Oe/s. The black and red colored traces are obtained during forward and backward sweepings, respectively.

1313


\end{widetext}


\begin{thebibliography}{10}

\bibitem{Fay} D. Fay and J. Appel, {\it Phys. Rev. B} \textbf{22}, 3173 (1980).

\bibitem{Julian} S. R. Julian, Physics \textbf{5}, 17 (2012).

\bibitem{Hattori} T. Hattori, Y. Ihara, Y. Nakai, K. Ishida, Y. Tada, S. Fujimoto, N. Kawakami, E. Osaki, K. Deguchi, N. K. Sato, and I. Satoh, Phys. Rev. Lett. \textbf{108}, 066403 (2012).
 	
\bibitem{Machida} K. Machida and T. Ohmi, Phys. Rev. Lett. \textbf{86}, 850 (2001).
	
\bibitem{Shopova} D. V. Shopova and D. I. Uzunov, Phys. Rev. B \textbf{72}, 024531 (2005).

\bibitem{Saxena} S. Saxena, P. Agarwal, K. Ahilan, F. M. Grosche, R. K. W. Haselwimmer, M. J. Steiner, E. Pugh, I. R. Walker, S. R. Julian, and P. Monthoux, Nature \textbf{406}, 587 (2000).

\bibitem{Aoki} D. Aoki, A. Huxley, E. Ressouche, D. Braithwaite, J. Flouquet, J. P. Brison, E. Lhotel, and C. Paulsen, Nature \textbf{413}, 613 (2001).

\bibitem{Huy} N. T. Huy, A. Gasparini, D. E. de Nijs, Y. Huang, J. C. P. Klaasse, T. Gortenmulder, A. de Visser, A. Hamann, T. Gorlach, and H. von Lohneysen, Phys. Rev. Lett. \textbf{99}, 067006 (2007).

\bibitem{Bergeret} F. S. Bergeret, A. F. Volkov, and K. B. Efetov, Phys. Rev. Lett. \textbf{86}, 4096  (2001).

\bibitem{Kadigrobov}  A. Kadigrobov, R. I. Shekhter, and M. Jonson, Europhys. Lett. \textbf{54}, 394 (2001).

\bibitem{Keizer} S. R. Keizer, S. T. B. Goennenwein, T. M. Klapwijk, G. X. Miao, G. Xiao, and A. Gupta, Nature \textbf{439}, 825 (2006).

\bibitem{Sigrist1991} M. Sigrist and K. Kazuo, Rev. Mod. Phys. \textbf{63}, 239 (1991).

\bibitem{Mackenzie2003} A. P. Mackenzie and Y. Maeno, Rev. Mod. Phys. \textbf{75}, 657 (2003).

\bibitem{Maeno2012} Y. Maeno, S. Kittaka, T. Nomura, S. Yonezawa, and  K. Ishida, J. Phys. Soc. Jpn. \textbf{81}, 011009 (2012).

\bibitem{Nelson2004} K. D. Nelson, Z. Q. Mao, Y. Maeno, and Y. Liu, Science \textbf{306}, 1151 (2004).

\bibitem{Kidwingira2006} F. Kidwingira, J. D. Strand, D. J. Van Harlingen, and Y. Maeno, Science \textbf{314}, 1267 (2006).

\bibitem{Bouhon2010} A. Bouhon and M. Sigrist, New J. Phys. \textbf{12}, 043031 (2010).

\bibitem{JJang} J. Jang, D. G. Ferguson, V. Vakaryuk, R. Budakian, S. B. Chung, P. M. Goldbart, and Y. Maeno, Science \textbf{331}, 186 (2011).

\bibitem{Bjrnsson} P. G. Bj\"{o}rnsson, Y. Maeno, M. E. Huber, and K. A. Moler, Phys. Rev. B \textbf{72}, 012504 (2005).

\bibitem{Kirtley} J. R. Kirtley, C. Kallin, C. W. Hicks, E. A. Kim, Y. Liu, K. A. Moler, Y. Maeno, and K. D. Nelson, Phys. Rev. B \textbf{76}, 014526 (2007).

\bibitem{Hicks} C. W. Hicks, J. R. Kirtley, T. M. Lippman, N. C. Koshnick, M. E. Huber, Y. Maeno, W. M. Yuhasz, M. B. Maple, and K. A. Moler, {\it Phys. Rev. B} \textbf{81}, 214501 (2010).

\bibitem{Moodera} J. S. Moodera and R. Meservey Phys. Rev. B \textbf{42}, 179 (1990).

  \bibitem{LeClair} P. LeClair, J. S. Moodera, J. Philip, and D. Heiman, Phys. Rev. Lett. \textbf{94} 037006 (2005).

\bibitem{Gong} X. X. Gong, H. X. Zhou, P. C. Xu, D. Yue, K. Zhu, X. F. Jin, H. Tian, G. J. Zhao, and T. Y. Chen, Chin. Phys. Lett. \textbf{32}, 067402 (2015).

\bibitem{Gong1} X. X. Gong, M. Kargarian, A. Stern, D. Yue, H. X. Zhou, X. F. Jin, V. M. Galitski, V. M. Yakovenko, and J. Xia, arXiv:1609.08538.

\bibitem{Sigrist1989} M. Sigrist, T. M. Rice, and K. Ueda, Phys. Rev. Lett. \textbf{63}, 1727 (1989).

\bibitem{supplementary_materials} See Supplemental Material at [URL will be inserted by publisher] for more data and discussions.

\bibitem{Barone} A. Barone, {\it Physics and Application of the Josephson Effect} John Wiley and Sons, Inc. (1982).

\end{thebibliography}
\end{document}